\documentclass{mn2e}

\newcommand{\etal}{{et al.}}
\newcommand{\kms}{\mbox{${\,\rm km~s}^{-1}$}\,}
\newcommand{\kev}{\,\mbox{${\,\rm keV}$\,}}
\newcommand{\ergs}{\mbox{${\,\rm erg~s}^{-1}$\,}}
\newcommand{\xmm}{\mbox{\sl XMM-Newton}}
\newcommand{\rosat}{\mbox{\sl ROSAT}}
\newcommand{\einstein}{\mbox{\sl EINSTEIN}}
\newcommand{\chandra}{\mbox{\sl Chandra}}
\newcommand{\hi}{\mbox{H{\small I}}}
\newcommand{\halpha}{\mbox{H{$\alpha$}}}

\title[\xmm\  EPIC Observations of the Starburst Galaxy M82] {First Look
\xmm\  EPIC Observations of the Prototypical Starburst Galaxy M82}

\author[ I.R. Stevens, A.M. Read, J. Bravo-Guerrero]
{Ian R. Stevens$^{1}$, Andrew M. Read$^{1}$, Jimena Bravo-Guerrero$^{1}$\\  
$^{1}$ School of Physics and Astronomy, University of Birmingham, 
Edgbaston, Birmingham, B15 2TT, UK\\
(E-mail: irs@star.sr.bham.ac.uk, amr@star.sr.bham.ac.uk, 
jbg@star.sr.bham.ac.uk)}  

\date{  Accepted ..............................; 
        Received ..............................; 
in original form ..............................}

\begin{document}
\maketitle

\begin{abstract} 

We present initial \xmm\  EPIC observations of the prototypical starburst
galaxy M82. The superwind is seen to extend continuously from the
starburst region to the X-ray emission associated with the \halpha\ \lq
cap\rq. We also find evidence for a ridge feature, probably associated
with the superwind interacting with an \hi\  streamer. Narrow band
images, centred on individual X-ray lines, show differences in
morphology, with higher energy lines being less spatially extended, and
a systematic shift with energy in the region of peak emission. Spectral
fits with two thermal and one power-law component provide a good fit to
the spectra of the nuclear and inner wind regions, indicating a
multiphase superwind. We discuss the implications of these observations
on our understanding of superwinds.

\end{abstract}


\section{Introduction}

Starburst events in galaxies represent an important phase in the
evolution of galaxies and ultimately in the generation of cosmic
structure (Heckman 2002). Starbursts affect galactic structure and
evolution by heating and metal-enriching the ISM via material outflowing
from the starburst, and if the event is strong enough, removing some of
the ISM via a galactic superwind. Indeed, if the starburst event is
strong enough (or the galaxy small enough) a starburst can remove the
majority of the ISM, with major implications for future star-formation
(Mac Low \& Ferrara 1999). Superwinds can have a major impact on their
environments, enriching the IGM, suppressing galaxy formation and so
on. The impact of these events may have been particularly important at
high redshift (Mori, Ferrara \& Madau 2002).

In order to study in detail the impact of superwinds on galactic
evolution and the IGM it is necessary to look at nearby objects, where
we have better spatial resolution and photon statistics. Nearby
superwind galaxies, such as NGC\,253, NGC\,1569 and M82, are key objects
and because superwinds are driven by hot shocked gas, originating from
supernovae or fast stellar winds, then X-rays are probably the most
important waveband to study superwinds.  While \chandra\  is a superb
instrument for disentangling point sources in galaxies, the large
collecting area of \xmm\  makes it  better for studying extended low surface
brightness features. Some important \chandra\  and \xmm\  results on
starbursts and superwinds have already been presented (eg Pietsch \etal\
2001; Strickland \etal\  2000; Martin, Kobulnicky \& Heckman 2002). In
this paper we present the first \xmm\  European Photon Imaging Camera
(EPIC) results for M82. The first M82 results from the \xmm\  Reflection
Grating Spectrograph (RGS) have been presented in Read \& Stevens
(2002).

\begin{figure*}
\vspace{9cm}
\includegraphics{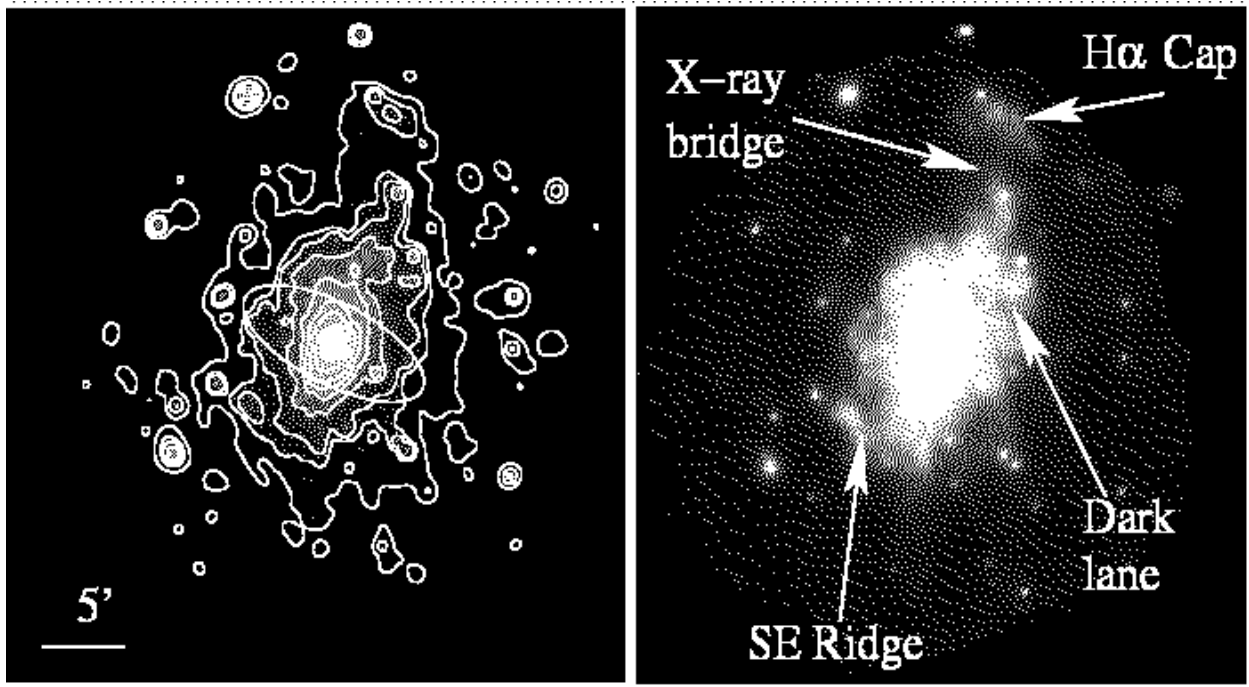} 
\caption{Left: The \xmm\  view of M82. An adaptively smoothed and
exposure-corrected image of the merged data from the MOS1, MOS2 and PN
instruments in the $0.2-10\kev$ waveband. The extent of the optical
galaxy is shown by the ellipse ($11.2'\times 4.3'$). Contours are
shown to highlight the low surface brightness emission, and increase
by factors of 2.  Right: The three-colour EPIC image of M82. The 
$0.2-0.5\kev$, $0.5-0.9\kev$ and $0.9-2.0\kev$ bands are shown by red, 
green and blue respectively. The main morphological features discussed 
in the paper are indicated.}
\label{fig1}
\end{figure*}

M82 is a much studied, nearby starburst and superwind galaxy
($D=3.63$~Mpc, Freedman \etal\  1994, so that $1'=1.06$~kpc). M82 is
very IR luminous (Rieke \etal\  1980), contains a substantial
population of young supernova remnants and luminous H{\small II}
regions (Pedlar \etal\ 1999), and many luminous super star clusters
(de Grijs, O'Connell \& Gallagher 2001), all indicating a strong and
ongoing starburst, probably triggered by a close encounter with M81.

As regards the M82 superwind, extended, extraplanar emission is seen at
several wavelengths. \halpha\ observations show a filamentary structure
consistent with a roughly conical bipolar outflow (Shopbell \&
Bland-Hawthorn 1998), with implied gas velocities along the cone surface
of $\sim 600\kms$. The implied \halpha\ morphology is roughly
cylindrical for a height above the disk of $z<350$~pc, flaring out at
larger $z$ with an opening angle of $30^\circ$ (McKeith \etal\ 1995). An
additional structure, termed the \halpha\ \lq cap\rq\ lies at a distance
of $\sim 11-12$~kpc to the NW of M82 and is probably associated with the
superwind (Devine \& Bally 1999).  

M82 has been observed by virtually all X-ray satellites and only a
selective overview is given here. \einstein\ first detected extended
X-ray emission associated with M82, extending for several kpc above
the plane and being well correlated with the optical filamentary
structure (Watson, Stanger \& Griffiths 1984; Fabbiano
1988). Strickland, Ponman \& Stevens (1997), using \rosat, also
studied the extraplanar emission, extending to $z\sim 6$~kpc, with
asymmetric morphology and brightness. Fitting spectra to subregions in
the wind, they found that the fitted temperature of superwind emission
drops from $kT\sim 0.6\kev$ near the nucleus to $\sim 0.4\kev$ in the
outer wind. Strickland \etal\ (1997) concluded that the superwind
X-ray emission was most likely coming from shocked clouds within the
superwind rather than the superwind itself. Lehnert, Heckman \& Weaver
(1999) discussed an X-ray feature seen by \rosat\ lying about $11'$ N
of M82, coincident with the \halpha\ \lq cap\rq\ emission, and
concluded that this emission was likely produced by shock heating, as
the superwind encounters a cloud in the halo of M82. This highlights a
major unresolved issue -- the origin of the X-ray emission from the
superwind -- is it due to the wind itself or shocked dense clumps
within the wind, and how much of the superwind material is from the
starburst and how much is entrained by the outflow?  As highlighted by
Strickland \& Stevens (2000) understanding the origin of the X-ray
emission is important to understanding the mass and energy contained
in the wind, which in turn is crucial to understanding the impact of
superwinds on their host galaxies and environments. High resolution
\chandra\ observations of M82 have provided insight into the behaviour
of point sources in the central region. The brightest source is
located away from the galaxy centre and may be an intermediate mass
black-hole or a beamed X-ray binary (Kaaret \etal\ 2001).

Here we focus on an initial look at the morphology and spectra of the
superwind. A more detailed analysis will be presented in a later paper.

\begin{figure*}
\vspace{7.0cm} 
\includegraphics{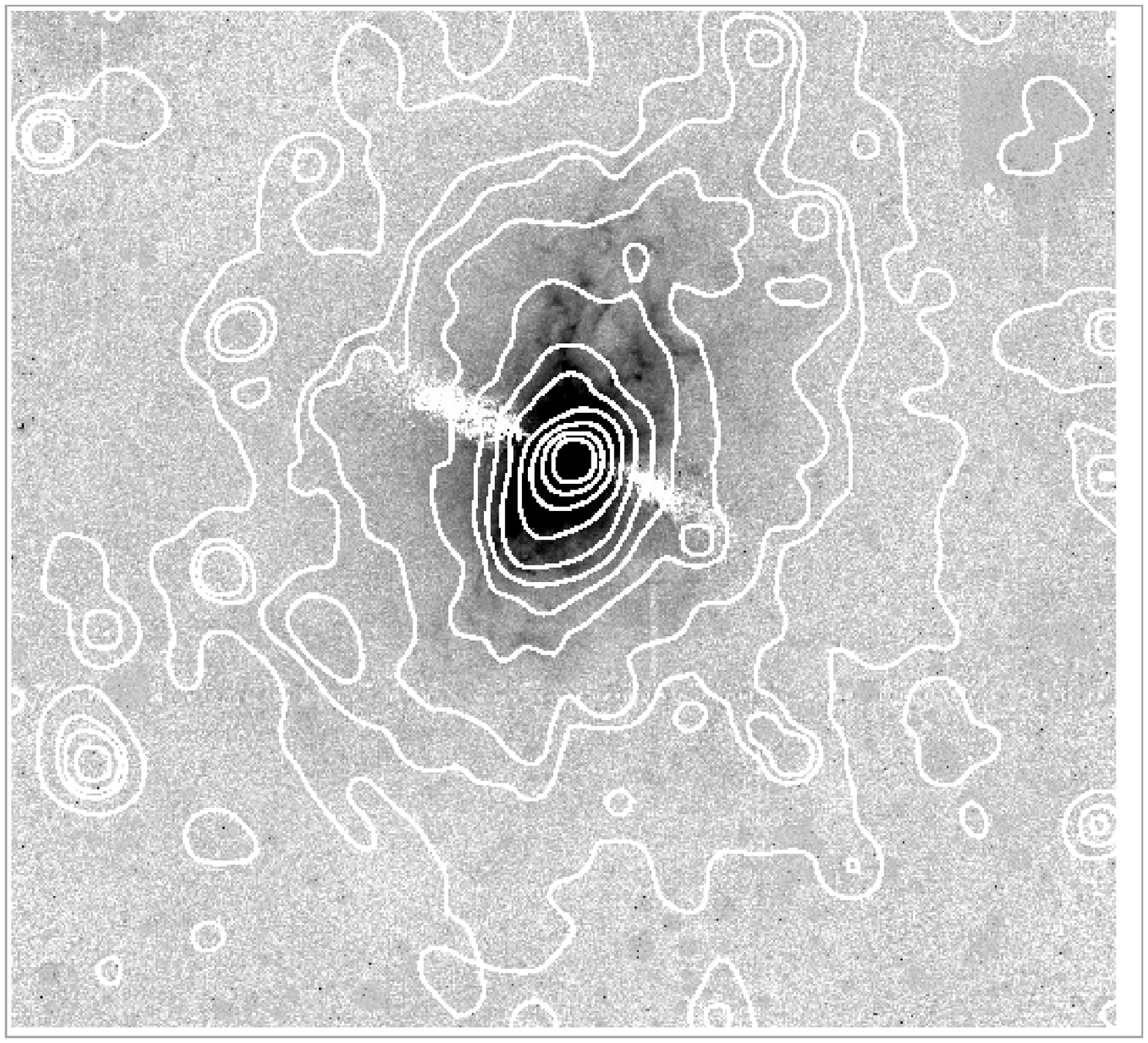}   
\includegraphics{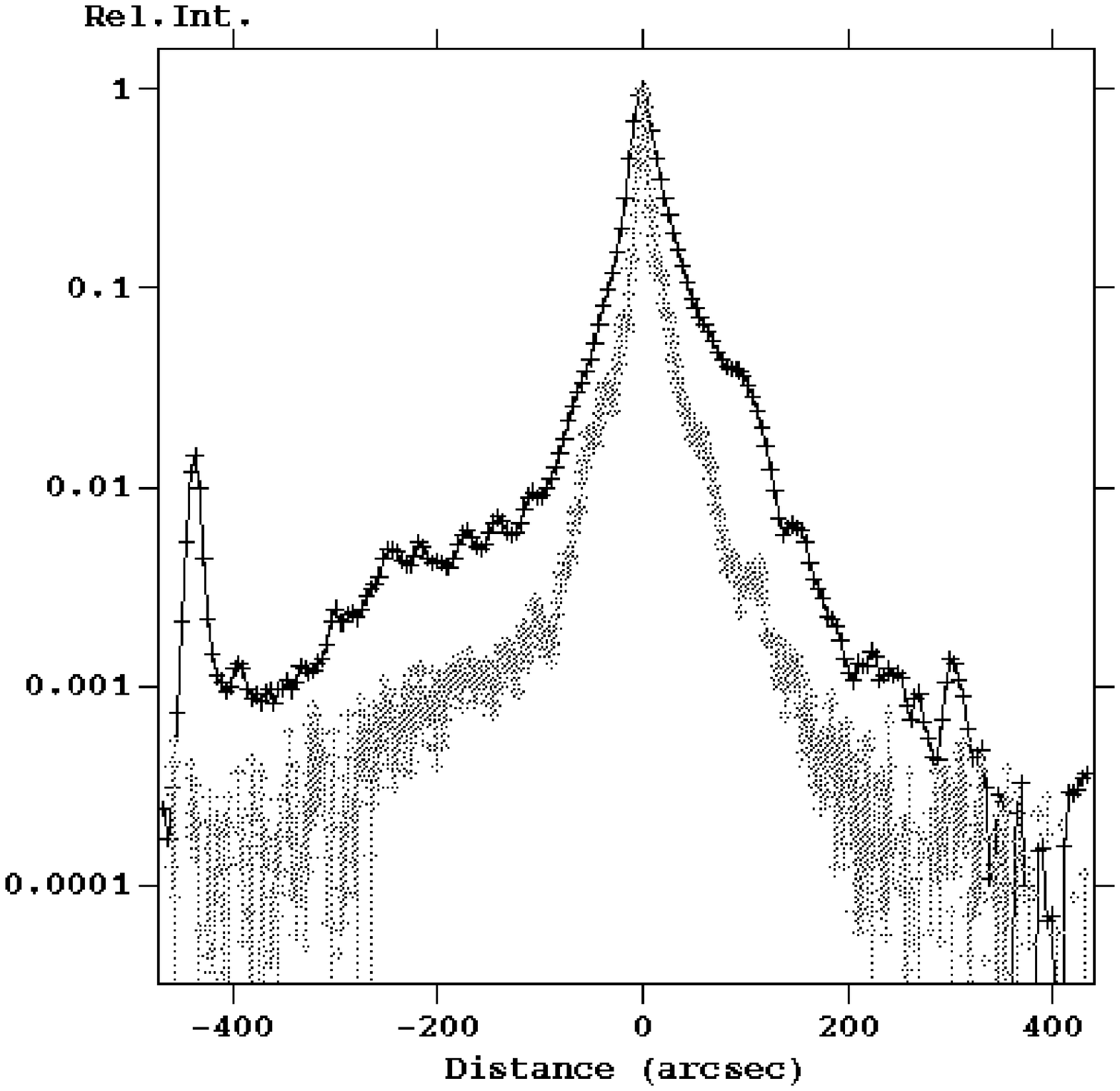}   
\caption{Left: The M82 X-ray/\halpha\ overlay. The X-ray contours from
the combined EPIC MOS and PN image are superimposed on a \halpha\
image. The image size is approximately $17'$ square. Right: The
relative intensities (normalised to the peak values) of the X-ray
(black) and \halpha\ emission (grey) as a function of distance from
the M82 nucleus in a slice running northeast (left) to southwest
(right) at a position angle of 25$^{\circ}$. The X-ray source at
$\alpha[2000.0]=09^{\rm h}55^{\rm m}14^{\rm s}$
$\delta[2000.0]=+69^{\circ}47' 32'' $) is visible to the very left of
the plot.}
\label{fig2}
\end{figure*}

\section{The \xmm\ Observations of M82}

M82 was observed with \xmm\ on 2001 May 6 (revolution 258) for
30.9ksec, with the EPIC MOS and PN cameras in full frame mode with the
medium filter. The following are preliminary results from the EPIC
MOS1, MOS2 and PN data. Details of \xmm\ and the EPIC
MOS and PN instruments can be found in Jansen \etal\ (2001), Turner
\etal\ (2001) and Str\"uder \etal\ (2001) respectively.

The data from the 3 EPIC instruments have been processed with standard
procedures in the \xmm\ SAS (Science Analysis System V5.2). Periods of
high background were filtered out (leaving 30.0ksec of usable
data). Only events with a pattern number $\leq 12$ and flag selection
for good events were retained, and photons with an energy in the range
$0.2-10\kev$.  For the PN detector, \lq out-of-time\rq\ events
were screened out. The pipeline processing has identified over 330
point sources within the EPIC field, though many are spurious and
associated with the \lq\lq out of time\rq\rq\ events. We defer a
detailed study of the M82 point source population and the associated
field to a later paper.

\section{Results: Imaging the Superwind}

In Fig.~\ref{fig1} (left panel) we show a combined MOS1, MOS2 and PN
image of M82. The data for each instrument has been corrected for
periods of high background and out-of-time events, mosaiced together
and adaptively smoothed. In addition to the large number of point
sources in the field, the main features to note are the bright nuclear
region and the extended extraplanar emission, predominantly bipolar in
nature. The superwind emission is asymmetric, extending to a greater
distance in the N ($\sim 13'$) than the S ($7'$). For the northern
wind the emission extends nearly to the edge of the field of view. The
emission is also clearly highly structured.

There are a few more additional features to note; the enhanced emission
in the N, coincident with the \halpha\ \lq cap\rq,
and a clear X-ray bridge connecting this emission to the main superwind
emission. The X-ray emission associated with the \halpha\ \lq cap\rq\
also seems to be structured and contains a point source (which could
well be a background object). As had been suspected, this X-ray emission
is now clearly seen to be part of the superwind.

In Fig.~\ref{fig1} (right) we show a three colour image of the emission
from M82. The 3 colour bands (red: $0.2-0.5\kev$, green: $0.5-0.9\kev$,
blue: $0.9-2.0\kev$) have been chosen to highlight the softer superwind
emission. In addition to some harder point sources in the field, much of
the superwind emission is of a fairly consistent colour, implying little
temperature change outside the inner region. There does appear to be a
softening towards the edge of the superwind. As can be seen from the
spectral fits discussed later we see a reduction in both the fitted
temperatures and in the level of absorption suffered. Both these effects
are responsible for the softening of the wind emission.

The X-ray emission associated with the \halpha\ \lq cap\rq\ is
concentrated in the $0.5-0.9\kev$ (green) waveband.  Another
interesting morphological feature is a ridge of emission in the SE
portion of the wind (and which is indicated in
Fig.~\ref{fig1}). Similar ridge like features in the X-ray emission
from superwinds have been seen in NGC\,253 for example (Strickland
\etal\ 2000). In the N wind there is a region with a deficit of
emission, which we term the \lq dark lane\rq, which is well correlated
with \hi\ emission (see later).

In Fig.~\ref{fig2} we show X-ray contours from the combined EPIC MOS1,
MOS2 and PN image, superimposed on a \halpha\ image of M82. Note that
this \halpha\ image does not extend out to the \halpha\ \lq
cap\rq. There is a good general tie-up in the superwind, between the
filamentary \halpha\ and the X-ray emission. In the northern wind
there are indications of the X-ray emission more closely following the
filamentary structure, with \halpha\ emission associated with the
tongue of X-ray material connecting to the \halpha\ \lq cap\rq\
emission.

Fig.~\ref{fig2} also shows how the relative intensities of the X-ray
and the \halpha\ emission vary with distance, in a NW-to-SE slice (at
a position angle of 25$^{\circ}$) through the nucleus of the
galaxy. Both profiles have been background-subtracted. Though the
\halpha\ emission appears more centrally concentrated than the X-ray,
several similarities in the profiles are evident. Both profiles show,
that within the inner wind (up to $100''$) the southern side is
brighter. This is likely due to the fact that,
due to the orientation of the system, we are able to see the southern
side of the inner wind far more clearly than the obscured northern
side. Beyond this however, the situation reverses; both the X-ray and
the \halpha\ profiles show a flattening of the intensity distribution
to the N, whereas to the S, both profiles fall more rapidly,
and disappear into the background. The northern wind is larger and
extends further than the southern wind.

\begin{figure}
\vspace{8.2cm} 
\includegraphics{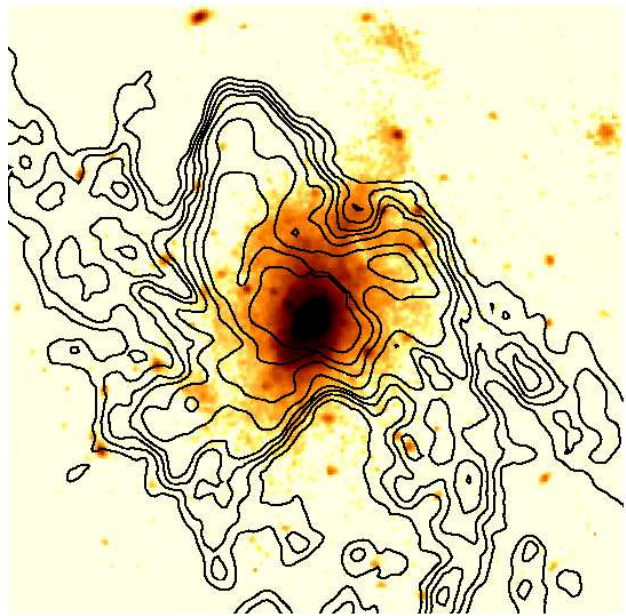}
\caption{Contours of \hi\ emission (constructed using the \hi\ data of
Yun \etal\ 1994) superimposed on the combined EPIC \xmm\ image of M82.} 
\label{fig3}
\end{figure}

There are some striking correlations (and anti-correlations) between the
X-ray emission and \hi\ features.  Contours of \hi\ emission,
constructed using the \hi\ data of Yun, Ho \& Lo (1994), are shown
superimposed on the \xmm\ image of M82 in Fig.~\ref{fig3}. An apparent
`hole' in the \hi\ emission is visible to the NW, coincident with the
northern X-ray wind. It is likely that the hot X-ray wind, in travelling
out into the IGM, has blown this hole in the intervening cold neutral
material. Note though, that the \hi\ features have masses comparable to
those implied for the superwind, and so could provide a substantial
obstacle for the outflow (Yun \etal\ 1993). The large \hi\ feature to
the east of the northern wind appears to be sharply bordered by X-ray
emission from the superwind, and may be collimating the flow. The 'dark
lane' in the X-ray emission in the N wind seems also to be associated
with enhanced \hi\ emission. The \lq dark lane\rq\ structure seen in the
X-ray image could be due to either absorption by foreground \hi\
emission or due to an interaction is not clear. The fact that we do not
see a hardening of the spectra in this region in the 3 colour diagram
(Fig.~\ref{fig1}) suggests that it is not simply foreground absorption.

The southern wind does not show such an obvious \hi\ `hole' as the
northern wind, but the X-ray ridge is interesting, lying at the
southern edge of the large \hi\ streamer extending to the east. The
\hi\ streamers to the southwest (in the direction of M81) do not show
any significantly enhanced X-ray emission. Also, the \halpha\ \lq
cap\rq\ does not show any associated \hi\ emission, and hence has been
assumed to be due to an ionized cloud.

\begin{figure}
\vspace{7.4cm} 
\includegraphics{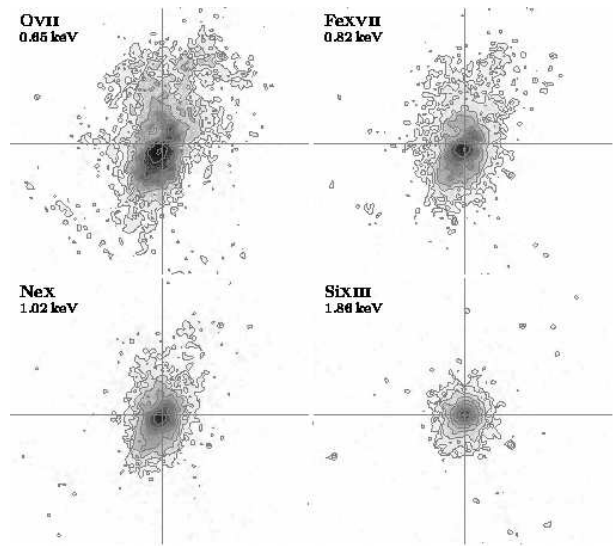}
\caption{Narrow band images, centred on the following lines:
O{\small VIII} (top left), Fe{\small XVII} (0.82\kev) (top right),
Ne{\small X} (bottom left) and Si{\small XIII} (bottom right). Each
image is $12'$ to a side, and the contours are at the same
levels in each image and increase by factors of two. The axes are
located at the centroid of the  Si{\small XIII} emission and show the
systematic shift in peak emission.}  
\label{fig4}
\end{figure}

\begin{table*}
\caption{M82 spectral fits for the nuclear region, inner wind region and
the \halpha\ \lq cap\rq\ region. The 90\% confidence intervals are
quoted, as is the emission fraction from each spectral component. The
abundance value applies to both the {\it mekal} components for the
nuclear and inner wind region. The model component normalisations are
also given -- for the {\it mekal} models the normalisation is
$(10^{-14}/4\pi D^2) \int n_e n_h dV$, with $V$ the volume and $D$ the
distance, while for the power-law models the units for the normalisation 
are photons keV$^{-1}$ cm$^{-2}$ s$^{-1}$ at 1 keV). The luminosity is
the absorbed broad band luminosity of the region, assuming $D=3.63$~Mpc.} 
\begin{tabular}{ccccccccc} \hline
Region & Model & $N_{H}$ & $kT/\Gamma$ & $Z$ & Normalisation & $L_X$ & 
Fraction &$\chi^2_\nu$ \\ 
& Component & (10$^{22}$ cm$^{-2}$) & (\kev)  & ($Z_\odot$) & & (\ergs)  
&  (\%) & (d.o.f) \\ \hline
Nuclear & {\it mekal} & $0.17_{0.16}^{0.18}$ & $0.54_{0.53}^{0.55}$ &
$0.30_{0.29}^{0.32}$ & $3.08_{2.71}^{3.40}\times 10^{-3}$ &
$3.70 \times 10^{40}$ & 7 & 1.41 \\ 
Region & {\it mekal} & $1.46_{1.43}^{1.48}$ & $0.90_{0.89}^{0.91}$  & &
$4.16_{2.97}^{4.27}\times 10^{-2}$  & & 22 & (2136)\\
& Power-law & $8.14_{7.77}^{8.63}$ & $1.81_{1.75}^{1.87}$ &-- &
$7.97_{7.16}^{9.15}\times 10^{-3}$ && 71 & \\
\hline
Inner Wind  & {\it mekal} & $0.07_{0.06}^{0.08}$ & $0.37_{0.36}^{0.38}$ &
$0.30_{0.28}^{0.33}$ & $3.13_{2.8}^{3.5}\times 10^{-3}$ & $1.08 \times
10^{40}$ & 27 & 2.15\smallskip \\ 
Region  & {\it mekal} & $0.89_{0.86}^{91}$ & $0.74_{0.73}^{0.76}$ & &
$1.55_{1.45}^{1.63} \times 10^{-2}$ & & 36 & (1069) \smallskip \\
 & Power-law & $8.30_{7.65}^{8.88}$ & $2.40_{2.30}^{2.51}$ & -- &
$3.41_{2.51}^{4.79}\times 10^{-3}$ & & 37 & \\ \hline
\halpha\ \lq Cap\rq\ & {\it mekal} & $0.04_{0.00}^{0.18}$ &
$0.65_{0.62}^{0.69}$ & 
$1.0$ (fixed) & $3.28_{2.64}^{5.03}\times 10^{-5}$ & $1.30\times
10^{38}$ & -- & 1.32 (162)\\ \hline 
\end{tabular}
\label{tab1}
\end{table*}

The large effective area of the EPIC instruments on \xmm\ allows the
generation of narrow band images, centred on particular lines, which
gives further information on the thermal and (potentially) the abundance
structure of superwinds.  In Fig.~\ref{fig4} we show 4 such images, with
each image centred in energy space on 4 lines, from low energy lines
(O{\small VIII}, 0.65\kev) up to higher energy lines (Si{\small XIII},
1.85\kev). The differences in
morphology are marked, with the spatial extent of the higher energy
lines being much smaller than the lower energy lines.  There is also a
systematic shift in the peak of the emission, with the emission peak of
the lower energy lines being shifted to the S and SE by up to
$\sim0.5'$. The orientation of the M82 disk is such that the NW edge of
the disk lies nearest to us (McKeith \etal\ 1995). As such, we are able
to see the central regions of the X-ray outflow much more easily in the
S than in the N. The energy shift in peak emission is very
likely due to a combination of two effects $-$ the central plume of the
very inner starburst wind (visible only in the south) above (i.e. to the
SE of) the nucleus is visible in the lower energy band images, whereas
the actual starburst nucleus is completely absorbed by the intervening
disk of M82. As we move towards higher energies, the contribution of the
central plume of the very inner wind emission decreases, whilst the
starburst nucleus becomes progressively less absorbed, and more
visible. An almost identical situation is seen in the other famous nearby
starburst galaxy NGC\,253 (Pietsch \etal\ 2001).  Note also, that the
X-ray spatial structure in the central regions of M82 has also been
studied by analysing the dispersion and cross-dispersion profiles of the
X-ray spectral lines observed with the RGS (Read \& Stevens 2002). These
results agree very well with the results presented here.

\section{Results: X-ray Spectra}

Previous X-ray spectral analyses of M82 have shown that multiple
spectral components are needed (for instance, Moran \& Lehnert 1997;
Ptak \etal\ 1997), indicating the presence of
multi-phase X-ray emitting material. In the light of these results we
adopt a spectral model with two thermal and one power-law components
(though we do note the ambiguities in results and physical
interpretation that can result from such models - see for example,
Dahlem \etal\ 2000). In addition, we find that one and two component
models give substantially inferior fits.

\begin{figure}
\vspace{17.25cm}
\includegraphics{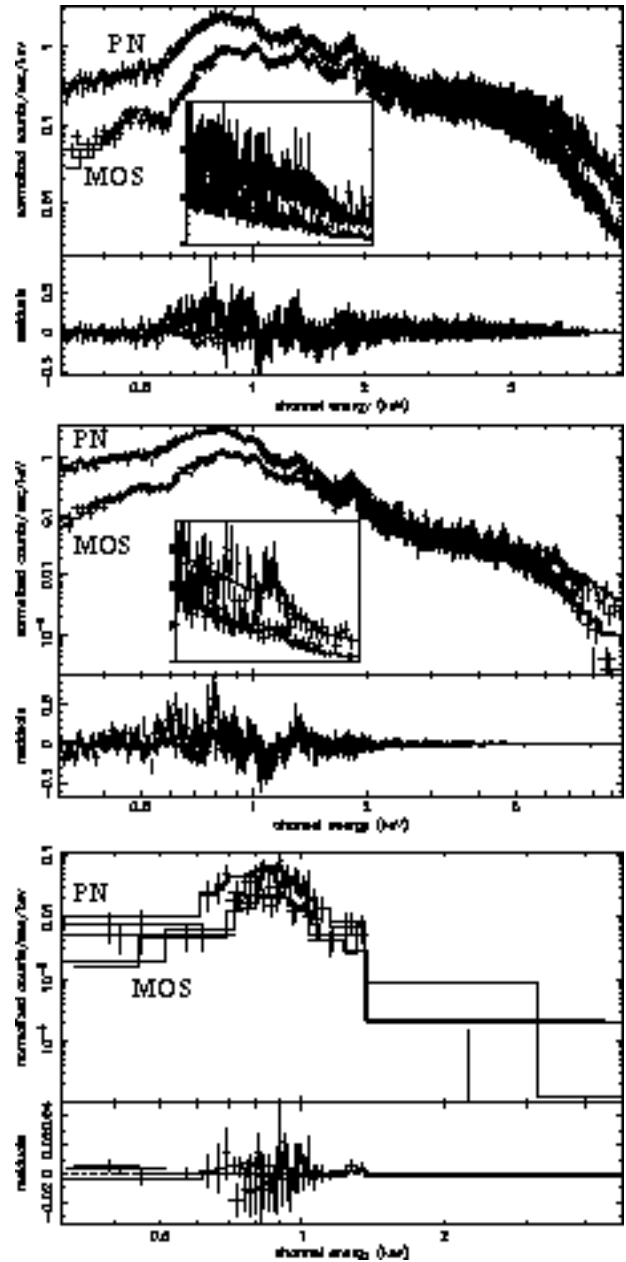}
\caption{The combined \xmm\ PN, MOS1 and MOS2 0.3--10.0\kev spectra of
the nuclear region (top), inner wind region (middle) and H$\alpha$ \lq cap\rq\ 
region (bottom), showing the best fit model and residuals - see
Table~\ref{tab1} for details of the fits. The insets in the top two
panels are blow-ups of the 5--8\kev spectra for the powerlaw+Gaussian
fit (Sect.~4)}  
\label{fig5}
\end{figure}

In this first EPIC paper we present a limited number of results for a
few regions of M82 and its superwind. More comprehensive results,
dealing with the spectral variations in the superwind are deferred to a
later paper. We have extracted spectra from all instruments, defining
two concentric source regions, a nuclear region of radius $\sim 0.5'$
and an annular inner wind region around the nuclear region, extending
from an inner radius of $0.5'$ to an outer radius of $1.3'$. The
background has been taken from a circular region reasonably close to
these regions but away from the diffuse emission.  We have used a three
component model to fit the PN and MOS1 and MOS2 spectra simultaneously,
two thermal ({\it mekal}) components with the abundance parameter tied
together and one power-law component, with each of the 3 components
having an independent column. The physical justification for this model
is that in the nuclear region, in addition to hot multi-phase superwind
gas, we expect harder emission from the point source population and also
possibly inverse Compton emission (Moran \& Lehnert 1997). In the inner
wind region, again in addition to the hot gas, we expect to detect
emission from unresolved point sources (background AGN as well as
sources associated with M82). The best-fit model results, absorbed
luminosities etc of each component are given in Table~\ref{tab1} and
shown in Fig.~\ref{fig5}. The approximate number of counts in each of
the spectra are: nuclear region -- PN: 99000, each MOS: 47000; inner wind
region -- PN: 65000, each MOS: 28000 and $\halpha$ cap region -- PN:4650,
each MOS:1800.  The power-law component in both regions is highly
absorbed (though the values are not inconsistent with the \hi\ values
for absorption in front of SNRs in M82 quoted by Wills, Pedlar \& Muxlow
1998) and has a slope broadly comparable to that of X-ray binaries and
AGN. Both thermal components are relatively soft, with values broadly
comparable to those seen with previous instruments. The single
metallicity value is (apparently) reasonably well constrained at $\sim
0.3Z_\odot$, though the reliability of this must be questioned (see
below). The X-ray emission spectrum of the \halpha\ \lq cap\rq\ region
is also shown in Fig.~\ref{fig5}. On account of the relatively low count
statistics the spectra are fitted with a single temperature {\it mekal}
model (see Table~\ref{tab1}), with $kT\sim 0.6\kev$. No firm constraints
can be placed on the metallicity of this region.

From these fits we can estimate masses of the X-ray emitting gas in
these regions (assuming spherical symmetry). Summing the masses in the
two thermal components, we find wind masses of $1.5\times
10^7\eta^{1/2} M_\odot$ for the nuclear region and $1.7\times
10^7\eta^{1/2} M_\odot$ for the inner wind region, where $\eta$ is the
volume filling factor of the X-ray emitting material (see Strickland
\& Stevens 2000 for a discussion of $\eta$). These values are broadly
comparable with those of Strickland \etal\  (1997), given the different
assumed geometries.

We expect the thermal emission from the superwind to come from both
nuclear processed material expelled from the starburst and material
entrained into the superwind from the galactic halo. The measured wind
abundances can then, in principle, be an important constraint on
superwind ejection processes, and be used to estimate the level of heavy
element pollution of the IGM by the starburst.
We can investigate the abundances of the nuclear and inner
wind region by grouping elements into 3 groups: (1) He, C, and N
(abundance set to solar), (2) Fe group -- Fe and Ni (associated with SNe
I), and (3) $\alpha$ elements -- O, Ne, Mg, Si, S, Ar and Ca (associated
with SNeII). Fitting again with a two temperature plus power-law model,
we find that (with the abundances tied between the two models) that the
Fe group abundance is fitted to be $0.43_{0.40}^{0.48}Z_{(Fe,\odot)}$
and for the $\alpha$ group $0.37_{0.35}^{0.40}Z_{(\alpha,\odot)}$,
implying a slightly subsolar $\alpha$/Fe ratio. For the inner wind
region the corresponding values are $0.29_{0.26}^{0.31}Z_{(Fe,\odot)}$
and $0.54_{0.50}^{0.57}Z_{(\alpha,\odot)}$, implying a supersolar ratio
(cf. Martin \etal\  2002 results for NGC\,1569).

The presence of an Fe line at $\sim 6.5\kev$ in the M82 spectrum has
been suggested by {\sl BeppoSAX} observations (Cappi \etal\ 1999). To
verify this we have fitted a power-law+Gaussian model to the $5-10\kev$
spectra of the nuclear and inner wind regions. For the nuclear region,
we find a line at $6.61_{6.54}^{6.68}\kev$, width
$\sigma=0.15_{0.08}^{0.25}\kev$ and a line flux of
$2.2_{1.4}^{3.7}\times 10^{-5}$ ph cm$^{-2}$ s$^{-1}$. For the inner
wind region, we find a line at $6.43^{6.31}_{6.53}\kev$ and a line flux
of $6.8_{3.0}^{18.9}\times 10^{-6}$~ph~cm$^{-2}$~s$^{-1}$, with $\sigma$
poorly constrained. These values are broadly consistent with Cappi
\etal\ (1999).

\section{Summary and Conclusions}

In this paper we have presented initial results from \xmm\ EPIC
observations of M82, which show the extraplanar X-ray emission
associated with the superwind of M82 in greater clarity than has been
seen before. The main features are that the superwind emission extends
out to a height of at least 14~kpc in the N and 7.5~kpc in the S and is
highly structured. The superwind is continuous out to the X-ray emission
associated with the \halpha\ \lq cap\rq. The X-ray and $\halpha$
profiles are similar, with the S wind brighter but overall less
extended. The are several interesting X-ray/\hi\ correlations. In the N
wind the superwind appears to have blown a hole in the \hi\ gas, as
revealed by a depression in the \hi\ emission coincident with a region
of X-ray emission. This penetration may explain why the N wind is more
extended. The is also some evidence of the superwind being collimated by
\hi\ gas. An X-ray ridge-like feature towards the SE edge of the wind
may be due to an interaction with an \hi\ streamer.

The X-ray spectra from the central regions of M82 can be fitted with a
three component model (2 thermal+power-law) with thermal components
with $kT= 0.5\kev$ and $0.9\kev$ (nuclear region), and $kT=0.4\kev$
and $0.7\kev$ (inner wind region). The emission is consistent with a
model with multiphase thermal emission from the superwind, along with
power-low emission from compact sources either in the galaxy or
background AGN.

In summary, we have presented only a short overview of the \xmm\
EPIC data on M82. More thorough work, which would include dealing with
the point source contamination of the superwind, is needed for a
detailed analysis of the superwind.  The X-ray emission from the
superwind is extremely complex and these observations have revealed
new evidence of interactions of the superwind with circumgalactic
material. Future work will concentrate on different aspects, such as
the bright central point sources, the spatial and thermal structure of
the superwind and the relationship to the observed \halpha\ emission
and the overall energetics of the starburst and the superwind.

\section*{Acknowledgements}

The \hi\ data used to produce Fig.~\ref{fig3} was kindly supplied to us
by M.S. Yun.

\end{document}